\documentclass[preprint,english,review]{elsarticle}
\usepackage{graphicx}
\usepackage{epstopdf}
\usepackage{color}
\usepackage{lineno}

\usepackage{babel}

\newcommand {\kgdtot} {160~}
\newcommand {\kgdefftot} {144~}
\newcommand {\jrundouze} {876.7~}
\newcommand {\jrundix} {122.3~}
\newcommand {\siglim} {1.0$\times$10$^{-7}$ pb~}
\newcommand {\sigmas} {80 GeV/c$^2$}
\newcommand{\kgd}{kg$\cdot$d~}
\newcommand{\kgdf}{kg$\cdot$d}

\begin{document}

\title{First results of the EDELWEISS-II WIMP search
         using Ge cryogenic detectors with interleaved electrodes}

\date{\today}

\author[]{The EDELWEISS Collaboration}

\author[irfu]{\\E.~Armengaud\corref{cor1}}
\author[ipnl]{C.~Augier}
\author[neel]{A.~Beno\^{\i}t}
\author[csnsm]{L.~Berg\'e}
\author[irfu]{O.~Besida}
\author[iek,fzk]{J.~Bl$\mbox{\"u}$mer}
\author[csnsm]{A.~Broniatowski}
\author[fzk]{A.~Chantelauze}
\author[csnsm]{M.~Chapellier}
\author[csnsm]{G.~Chardin}
\author[ipnl]{F.~Charlieux}
\author[csnsm]{S.~Collin}
\author[csnsm]{X.~Defay}
\author[ipnl]{M.~De~Jesus}
\author[ipnl]{P.~Di~Stefano\fnref{pa-distefano}}
\fntext[fn1]{Present address: 
Department of Physics, Queen's University, Kingston, ON, K7L 3N6, Canada}
\author[csnsm]{Y.~Dolgorouki}
\author[csnsm,irfu]{J.~Domange}
\author[csnsm]{L.~Dumoulin}
\author[fzk]{K.~Eitel}
\author[ipnl]{J.~Gascon\corref{cor2}}
\author[irfu]{G.~Gerbier}
\author[irfu]{M.~Gros}
\author[irfu]{M.~Hannawald}
\author[irfu]{S.~Herv\mbox{\'e}}
\author[ipnl]{A.~Juillard}
\author[fzk]{H.~Kluck}
\author[fzk]{V.~Kozlov}
\author[irfu]{R.~Lemrani}
\author[lsm]{P.~Loaiza}
\author[jinr]{A.~Lubashevskiy}
\author[csnsm]{S.~Marnieros}
\author[irfu]{X-F.~Navick}
\author[csnsm]{E.~Olivieri}
\author[iramis]{P.~Pari}
\author[irfu]{B.~Paul}
\author[jinr]{S.~Rozov}
\author[ipnl]{V.~Sanglard}
\author[ipnl]{S.~Scorza}
\author[jinr]{S.~Semikh}
\author[irfu]{A.S.~Torrento-Coello}
\author[ipnl]{L.~Vagneron}
\author[ipnl]{M-A.~Verdier}
\author[jinr]{E.~Yakushev}

% 1
\address[neel]{CNRS-N\'{e}el, 25 Avenue des Martyrs, 
38042 Grenoble cedex 9, France}
% 2
\address[csnsm]{CSNSM,
IN2P3-CNRS, Universit\'e Paris XI, bat 108, 91405 Orsay,  France}
% 3
\address[iek]{Karlsruhe Institute of Technology,
Institut f\"{u}r Experimentelle Kernphysik, Gaedestr. 1, 76128 Karlsruhe, Germany}
% 4
\address[fzk]{Karlsruhe Institute of Technology,
Institut f\"ur Kernphysik, Postfach 3640, 76021 Karlsruhe, Germany}
% 5
\address[iramis]{CEA, Centre d'Etudes Saclay, 
IRAMIS, 91191 Gif-Sur-Yvette Cedex, France}
% 6
\address[irfu]{CEA, Centre d'Etudes Saclay, IRFU, 
91191 Gif-Sur-Yvette Cedex, France}
% 7
\address[ipnl]{IPNL, Universit\'{e} de Lyon, Universit\'{e} Lyon 1, 
CNRS/IN2P3, 4 rue E. Fermi 69622 Villeurbanne cedex, France}
% 8
\address[jinr]{Laboratory of Nuclear Problems, JINR, Joliot-Curie 6, 
141980 Dubna, Moscow region, Russia}
% 9
\address[lsm]{Laboratoire Souterrain de Modane, CEA-CNRS, 
1125 route de Bardonn\`eche, 73500 Modane, France}

\cortext[cor1]{Corresponding author} 
\cortext[cor2]{Corresponding author}

\begin{abstract}

The EDELWEISS-II collaboration has performed a direct search for
WIMP dark matter with an array of ten  400~g heat-and-ionization cryogenic
detectors equipped with interleaved electrodes for the rejection
of near-surface events.
Six months of continuous operation at the Laboratoire Souterrain de Modane have been achieved.
The observation of one nuclear recoil candidate above 20 keV in an effective exposure of \kgdefftot \kgd is interpreted in terms
of limits on the cross-section of spin-independent interactions
of  WIMPs and nucleons.
A cross-section of \siglim  is excluded at
90\%CL for a WIMP mass of \sigmas.
This result demonstrates for the first time the very high background rejection
capabilities of these simple and robust detectors in an actual
WIMP search experiment.
\end{abstract}

\begin{keyword}
Dark Matter, Cryogenic Ge detectors, WIMP searches.
\PACS 95.35.+d \sep 14.80.Ly \sep 29.40.Wk \sep 98.80.Es
\end{keyword}

\maketitle

% \linenumbers

\section{Introduction}

The existence of Weakly Interacting Massive Particles (WIMPs) is a likely 
explanation for the various observations of a dark matter component from
the largest scales of the Universe to galactic scales~\cite{rev}. 
WIMPs are predicted by several extensions of the Standard Model of particle physics.
WIMPs distributed in the Milky Way halo may be detected through coherent, elastic
scattering on nuclei constituting a terrestrial detector~\cite{goodman}.
The expected nuclear recoils have a quasi-exponential energy distribution 
with typical energies of a few tens of keV.
Current searches~\cite{cdms,xenon,zeplin} set upper limits on the interaction rate
at the level of $\sim$10$^{-2}$ event per kg and per day (evt/kg/d).
Minimal extensions of the Standard Model with Supersymmetry where
the WIMP is the lightest neutralino predict a particularly interesting range
of parameters where their spin-independent scattering cross-section on 
nucleons lies between 10$^{-8}$ and 10$^{-10}$~pb,
corresponding to rates of 10$^{-3}$ to 10$^{-5}$ evt/kg/d.
As a consequence, experiments dedicated to the direct detection 
of WIMPs require large masses of detectors and long exposure times.
The radioactive background is the main obstacle to measure such extremely low rates.
Provision of passive rejection, such as the use of shieldings and radiopure
materials in a deep underground site, must be complemented by a
detector technology that enables a clear identification of
single nuclear recoils with respect to other types of interactions.

Cryogenic Germanium detectors~\cite{cdms,edw} constitute a leading technology
in direct detection of dark matter.
Event discrimination is provided by the comparison of the ionization signals to the
bolometric measurement of the deposited energy. 
A long-standing issue with these detectors is the reduction
of the charge collection efficiency for interactions occurring close to their surface, which can
impair significantly the discrimination between nuclear and electron recoils. 
The CDMS collaboration addresses it by adding a discrimination based on the time
structure of their athermal phonon signals~\cite{cdms}, 
obtaining with this the best published sensitivity for WIMPs of masses above $\sim$40 GeV/c$^2$.
Recently, the EDELWEISS collaboration deployed an alternative
solution based on a new detector design 
with interleaved electrodes~\cite{brinkbroni}, named ID in the following. 
In Ref.~\cite{id}, we demonstrated a rejection factor greater than 
10$^{4}$ for low-energy surface events induced by
electrons emitted from $^{210}$Pb source.
Combined with a comparable factor for the rejection of bulk electronic recoils, 
this technology, and the low-background environment achieved in the EDELWEISS-II
setup~\cite{edwsetup}, should enable to reach the required
sensitivity to probe WIMP-nucleon cross-sections well below 10$^{-8}$ pb. 
This new technology is also considered for application in more ambitious searches 
in the 10$^{-10}$~pb range such as in the EURECA project~\cite{eureca}.

The work presented in this letter confirms these results in the conditions of an actual 
WIMP search, and in particular demonstrates the detector reliability 
and efficiency over long periods of time.
We report on the results of a WIMP search carried out
over a period of six months
with an array of ten 400~g cryogenic germanium ID detectors.
These data are combined with a smaller set recorded during the
validation run of the first two ID 400~g detectors and interpreted in terms of limits
on the spin-independent WIMP-nucleon cross-section.
The reliability and efficiency of the ID technology will be assessed,
together with its prospect for a significant improvement of 
the sensitivity for WIMP detection.
These results will be compared to those of other leading WIMP searches.

\section{Experimental setup}

The experimental setup is described in detail in Ref.~\cite{edwsetup}.
It is located in the Laboratoire Souterrain de Modane (LSM).
The rock overburden of 4800 mwe reduces the cosmic muon flux to 4~$\mu$/m$^{2}$/day.
The flux of neutrons above 1~MeV is 10$^{-6}$ n/cm$^2$/s~\cite{lemrani}.
The cryostat housing the detectors is protected from the ambient $\gamma$-rays
by a 20~cm lead shield.
This is surrounded by a 50~cm thick polyethylene shield,
covered by a muon veto system with 98\% geometric efficiency 
for throughgoing muons.
The remaining rate of single nuclear recoils from energetic neutrons induced by muons
is less than $10^{-3}$ evts/kg/day
according to GEANT4 simulations of the experiment with its
shields and the active rejection of muon-tagged events~\cite{markus}.

The detectors are made from ten hyperpure germanium crystals 
(less than $10^{10}$ impurities per cm$^{3}$)  of cylindrical shapes 
with a diameter of 70 mm and a height of 20 mm. 
Five of these detectors have their edges bevelled at an angle of 45$^{o}$
and have a mass of 360~g. The mass of the other five detectors is 410~g. 
The detectors are in individual copper casings, stacked in towers of
two to three ID detectors.
During the entire data-taking periods, a dilution refrigerator maintains
the detectors at a stabilized temperature of 18~mK.

For each interaction in a detector, two types of signals are recorded:
an elevation of temperature and a charge signal.
The temperature increase for each event is measured using neutron transmutation
doped (NTD)-Ge  thermometric sensors glued on each detector,
as was the case in the previous EDELWEISS detectors~\cite{edw}.
The charge is collected by electrodes, the design of which
and polarization scheme are described in details in Ref.~\cite{id}.
Basically, the flat surfaces are covered with concentric 
aluminium ring electrodes of 2~mm pitch, and
biased at alternating potentials.
Each side has thus two sets of electrodes,
each connected to its next but one neighbor through ultra-sonically bonded wires.
The biases are chosen as to produce an axial field in the
bulk of the detector, while the field close to the surface
links two adjacent ring electrodes and is therefore approximately
parallel to the surface.

Bulk events are thus identified by the collection of electrons
and holes on the set of electrodes with the largest
difference of potential across the detector, called
{\em fiducial electrodes} in the following.
The two other electrodes act as a veto for surface events.
Typical biases for the fiducial and veto 
electrodes are $\pm 4$V  and $\mp 1.5$V respectively.
Two additional plain guard electrodes, typically biased at $\pm 1$V, 
cover the outer edges of the crystal. 
The fiducial volume of the detector is defined as the region
for which the charge is collected entirely by the two fiducial electrodes.

The heat and the six charge signals of each detector
are digitized continuously at a rate of 100 kHz, using a common clock
for all channels.
After online filtering, a threshold trigger is applied to
each of the heat channels. 
The value of that threshold is updated continuously within the data
acquisition on the basis of the noise level, aiming to keep the trigger 
rate below a fraction of a Hz.
For each event, the raw heat and ionization data of all detectors in 
the corresponding tower are stored to disk.
Triggers detected on more than one detector on the same tower
are tagged online as coincidences.
The data from detectors that did not participate to the trigger are
used to monitor systematically the baseline resolutions as a function
of time.

The muon veto has an independent read-out system.
Each of the 42 plastic scintillation modules is equipped with 8
photomultipliers, 4 added at each end.
A signal is recorded as a veto hit once a coincidence of the 2 ends of a 
module occurs within a 100~ns wide coincidence window. 
With a low trigger threshold deliberately chosen as to optimize
the muon veto efficiency, the muon veto rate of $\sim$0.2 events/sec is dominated 
by environmental background.
Each muon veto event is tagged with the corresponding time
on the common clock used for the cryogenic detector readout.
The coincidences between muon veto events and those
in the detector towers are identified offline by comparing the respective
time tags.

\section{Detector performance and fiducial acceptance}

After an initial cool-down of the cryostat in March 2009,
the bulk of the data presented here was recorded over the following
period of  six months from April to September.
An additional data set was recorded with
two detectors during an initial run performed between July and November 2008.

In the six-month running period, the data acquisition was running 
80\% of the time.
Half of the losses are accounted for by regular
maintenance operations and the other half by 
unscheduled stops.
The fraction of running time devoted to detector calibration
with $\gamma$ and neutron sources is 6\%.
Among the 70 read-out channels, 
only five were defective or too noisy for use:
four guards and one veto.
Extensive calibrations performed in 2008 have shown that 
a signal on one of the guards is almost
invariably accompanied by another one on
either the other guard or a veto electrode, or by
an imbalance in the charge collected on the 
fiducial electrodes.
Relying on this redundancy is sufficient to attain high
rejection factors even in the absence of a guard channel.
Three detectors with one deficient guard signal
are thus kept for the WIMP search.
One detector had one deficient guard and one deficient veto
electrode.
In the absence of evidence that this particular combination
of two missing channels can be efficiently compensated by 
the existing redundancies,
it was decided a priori not to use this detector for
WIMP searches.

The data were analyzed offline by two independent analysis chains. 
For each event, the amplitudes of the signals from the NTD-Ge
thermistors and the six electrode signals were determined
taking advantage of the synchronous digitization of all channels,
and using optimal filters adapted to their time-dependent
individual noise spectra.

The energy calibration of the ionization signal was performed with $^{133}$Ba $\gamma$-ray sources.
No gain variations were observed over the entire period, for a given
polarization setting.
The gain of the heat signal depends on the cryogenic conditions. 
It is measured as a function of time by monitoring the ratio of the heat 
and ionization signal for bulk $\gamma$-ray events.
The non-linearity of the heat gain as a function of energy is
measured using $\gamma$-ray calibrations.
The baseline resolution on the different heat channels 
varies between 0.6 and 2 keV FWHM, with an average of 1.2 keV.
The ionization baseline resolution for fiducial events 
varies between 0.8 and 1.2 keV FWHM, with an average of 0.9 keV.

The response of the ID detectors to nuclear recoils induced by neutron 
scattering was checked using an AmBe source.
For these events, the ionization yield $Q$, 
defined as the ratio of the charge signal to the recoil energy $E_R$
normalized to the response for $\gamma$-ray interactions,
follows a behavior consistent with the parametrization used
in previous EDELWEISS reports~\cite{edw}.
The nuclear recoil events are centered at $Q=0.16 \,E_R^{\;0.18}$,
where $E_R$ is in keV,
with a dispersion in agreement with the measured resolutions
of the ionization and heat channels~\cite{martineau}.
These calibrations were also used to verify that the acceptance
to low-energy nuclear recoil events is as predicted by the
known online trigger threshold and baseline resolutions
of the different signals.

\begin{figure}
\begin{center} \includegraphics[scale=0.7]{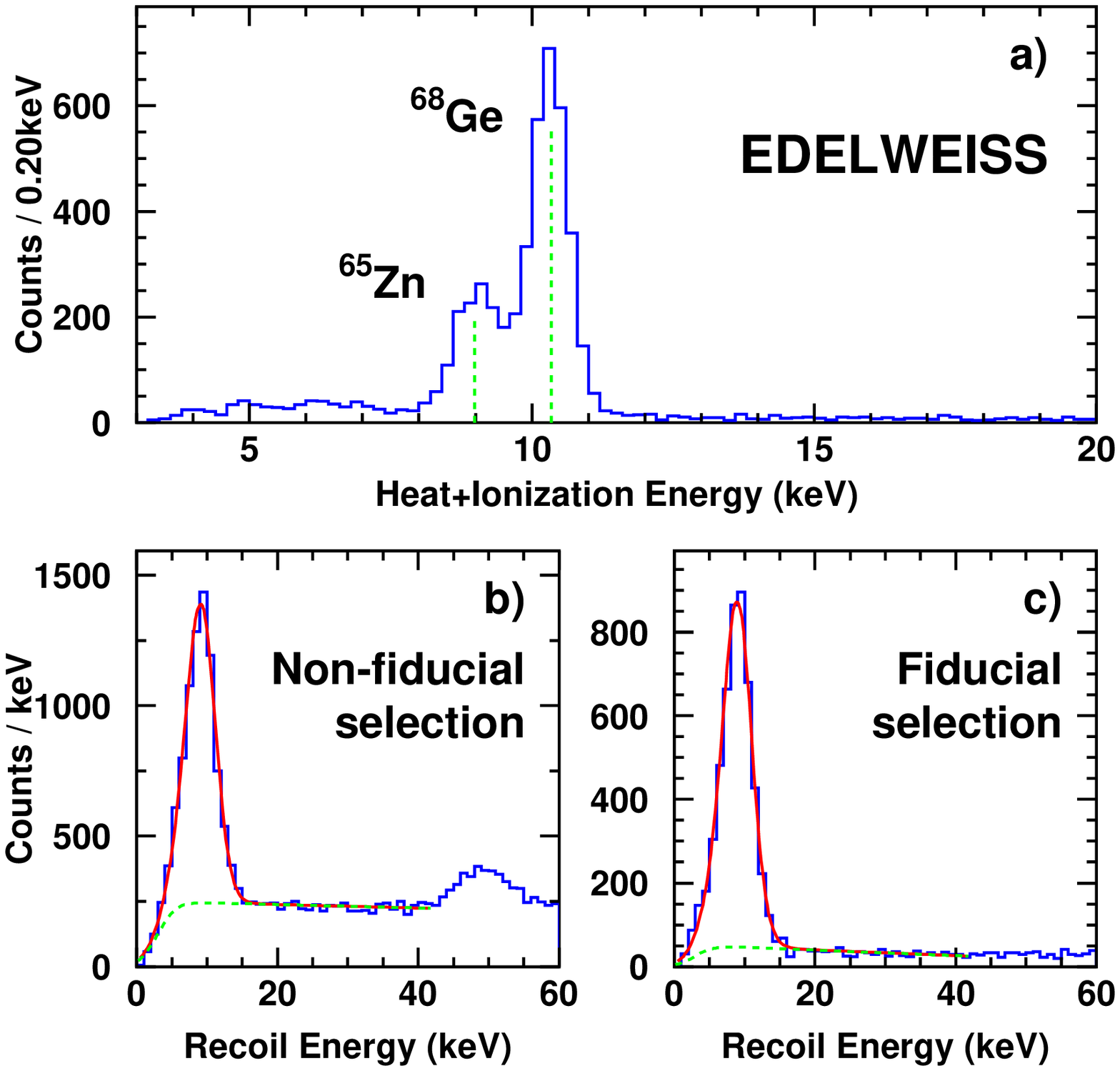} \end{center}
\caption{
Top: Energy spectrum of fiducial events recorded by the nine
detectors used in the WIMP search, using events with an online heat 
threshold below 3.5 keV.
The $\gamma$-ray energy is calculated from the sum of the heat and ionization signals 
weighted by their resolutions.
The two peaks at 8.98 and 10.34 keV are associated with the electron capture
decay of the cosmogenically activated isotopes $^{65}$Zn and $^{68}$Ge.
Bottom: Corresponding recoil energy spectra outside (b) and inside (c) the 
fiducial selection.
The 46 keV peak associated with $^{210}$Pb decays near the surface
of the detectors is clearly visible on (b) and strongly suppressed on (c).
Gaussian fits are drawn to guide the eyes.
The dashed lines† represent the acceptance function resulting from
3.5 keV threshold cuts applied on ionization and heat signals.
\label{fig:cosmo}}
\end{figure}

The identification of an event as fiducial requires that
each of the signals on veto and guard electrodes,
as well as the charge imbalance between the two
fiducial electrodes, be consistent with zero at 99\%CL, 
according to the resolution on these parameters.
The efficiency of those cuts, as well as the fraction of the
fiducial volume that they represent, was measured using
the low-energy peaks associated with the cosmogenic 
activation of germanium,
producing an uniform contamination by the isotopes
$^{65}$Zn and $^{68}$Ge.
The electron capture decays of these long-lived ($\sim$250 days)
isotopes result in energy deposition of 8.98 and 10.34 keV, 
respectively, as shown in Fig.~\ref{fig:cosmo}a.
Their yield throughout the detector volume can be measured
with an uniform efficiency by using recoil energy spectra.
Fig.~\ref{fig:cosmo}b and c show these distributions for, respectively,
events rejected and accepted by the fiducial cut, 
for all detectors.
The fraction of events in the peak that survive the fiducial
cuts is the fiducial efficiency.
This number fully takes into account the size of the electrostatic fiducial
volume, the effects of charge sharing between
electrodes due to their expansion and diffusion~\cite{broni-simu},
and the finite resolution on the electrode signals.
The study of the cosmogenic peaks is performed individually for each
detector.
The measured effective fiducial mass, averaged by the relative
exposure of each detector, is 166$\pm$6~g. 
This result is consistent with a fiducial
volume corresponding approximately to the cylinder defined by the last
fiducial electrodes, limited in height to $\sim$1~mm under the flat surfaces,
a value that is similar for detectors with and without
bevelled edges. 
A conservative value of 160~g will be used in the calculation of
the total exposure.

The average rate of events from cosmogenically activated isotopes 
in these recently installed detectors is approximately 4 per day.
To reject the several thousands that are expected in a six month exposure, 
events are accepted in the WIMP search if their ionization yield
is 3.72~$\sigma$ below $Q=1$, corresponding to a $\gamma$-ray
rejection factor of 99.99\%.
Given the measured detector resolutions, this requirement 
starts to affect the
efficiency in the 90\% nuclear recoil band below 20 keV. 
Another factor reducing the efficiency for nuclear recoils below 20 keV is the restriction
of the search to events with a fiducial ionization signal of at least 3 keV.
As a consequence, the following WIMP search is restricted {\em a priori} to recoil 
energies above 20 keV.
This value is well suited to test the rejection performance
of the detectors: in EDELWEISS-I, the event rate above this threshold 
in the nuclear recoil band was as high as 0.4 evt/kg/d~\cite{edw}. 
Lowering this threshold requires more studies to determine the acceptance in 
this region which depends on the detailed understanding of the noise 
on fiducial and non-fiducial ionization channels.

\section{Results and discussion}

In the six-month run, the total exposure for WIMP search of the nine detectors
is 1262 detector-days.
A time period selection is applied on the baseline resolutions and online heat
threshold on an hour-by-hour basis, in order to ensure that in each selected 
hour they are compatible with a nearly
full efficiency for nuclear recoils above 20 keV.
This requirement reduces the exposure by $\sim 20\%$. 
In addition, a cut is applied on the reduced $\chi^2$ of the 
pulse fit for heat and fiducial electrodes of each event. 
This is done in order to remove short periods of high-noise levels,
pile-up events and some populations of ionization-less 
events associated, for example, with radioactive decays in
the NTD-Ge sensor. 
The $\chi^2$ cut efficiency is controlled using the $\gamma$-ray background: 
its typical value is 97\%.
The deadtime associated with the online trigger algorithm is $\sim 1\%$ 
and an additional 3\% was lost due to DAQ problems.
Less than 1\% of exposure is lost due to the deadtime resulting
from the anti-coincidence requirement with the other bolometers
and with the muon veto within a 50~ms time window.

In the six-month run, after all cuts, the nine 
ID detectors have a total exposure of  \jrundouze detector-days.
The 2008 data with two of these nine ID detectors provides an additional exposure 
of \jrundix detector-days.
With the average fiducial mass of 160~g, this corresponds to a total of \kgdtot \kgdf. 
Finally, taking into account the 90\% C.L. region for nuclear recoils, 
this data set is equivalent to \kgdefftot \kgd with an acceptance 
of 92\% at 20 keV and 100\% at 23 keV.

\begin{figure}
\begin{center} \includegraphics[scale=0.7]{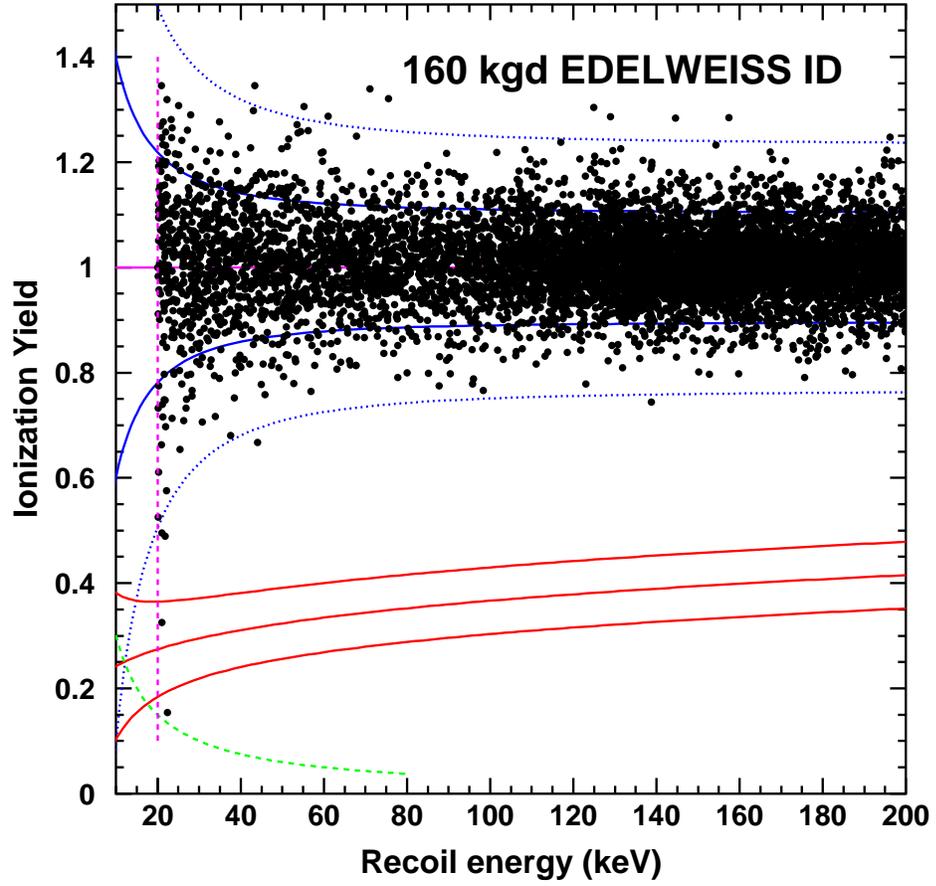} \end{center}
\caption{Ionization yield vs recoil energy of fiducial events recorded
by EDELWEISS-II in an exposure of \kgdtot \kgdf.
The WIMP search region is defined by recoil energies greater than 20 keV
(vertical dashed line). The 90\% acceptance nuclear and electron recoil
bands (full blue and red lines, respectively) are calculated using
the average detector resolutions. Also shown as dashed lines are
the 99.98\% acceptance band for the $\gamma$ (blue) and the
3 keV ionisation threshold (green). 
\label{fig:qplot}}
\end{figure}

Fig.~\ref{fig:qplot} shows the distribution of $Q$ as a function of recoil energy 
for the entire exposure.
The anti-coincidence requirement with the muon veto and with the other
detectors removes two events in the nuclear recoil band at 27 and 43 keV. 
The average rate outside the nuclear recoil band in the 20-100 keV
range is 0.16 event/keV/kg/d.

\begin{figure}
\begin{center} \includegraphics[scale=0.7]{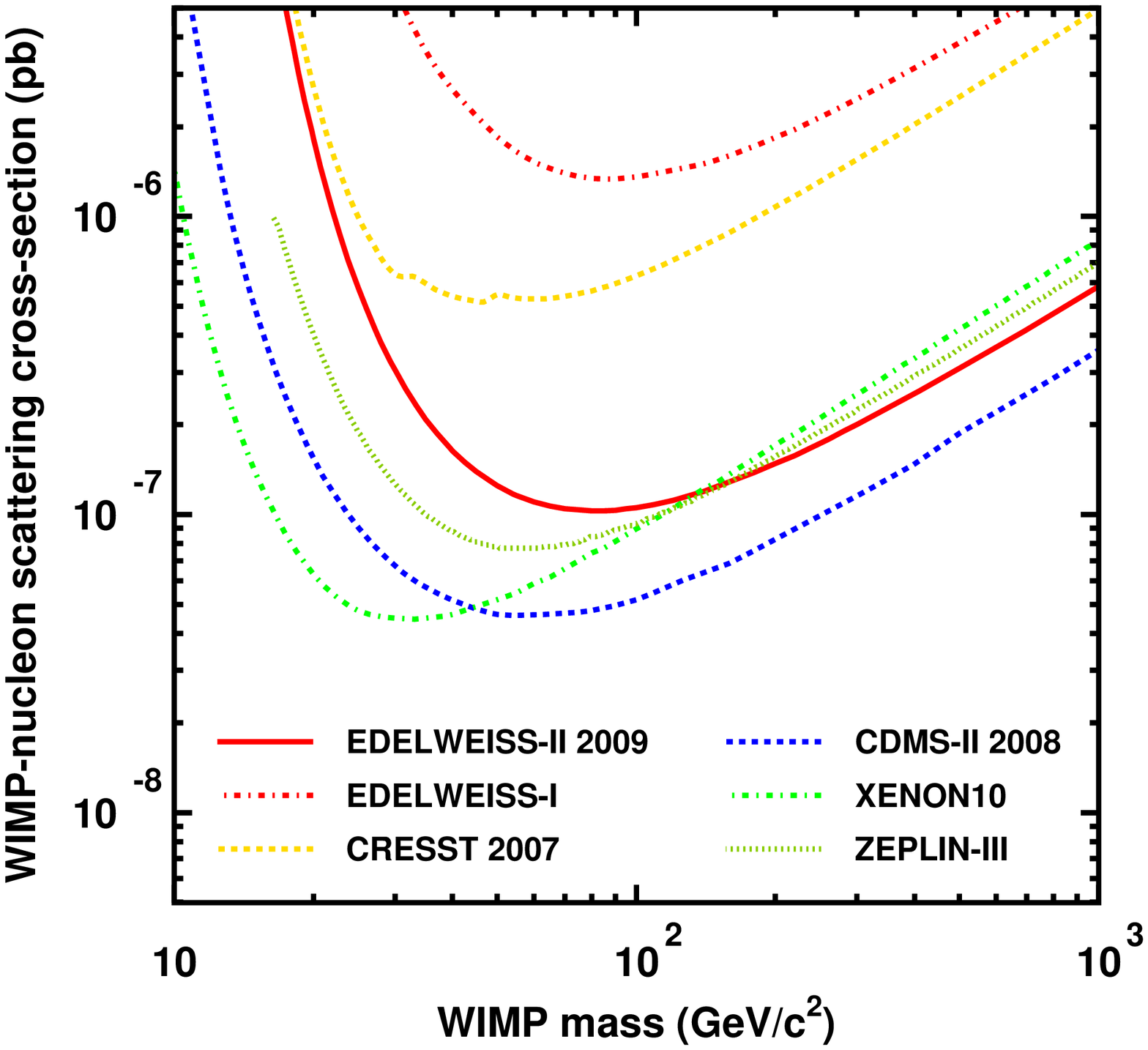} \end{center}
\caption{Limits on the cross-section for spin-independent scattering
of WIMPs on nucleons as a function of WIMP mass,
derived from the data presented in this letter,
together with limits from other direct WIMP searches.
The limits for CDMS, ZEPLIN, EDELWEISS-I and CRESST
are from Refs.~\protect{\cite{cdms,zeplin,edw,cresst}}, respectively.
The limits for XENON~\cite{xenon} do not include the modification
implied by the light yield measurements of Ref.~\protect{\cite{qxe}}.
\label{fig:limits}}
\end{figure}

One event is observed in the nuclear recoil band at 21.1 keV. 
This represents a factor $\sim$50 reduction relative to
the event rate above 20 keV measured in EDELWEISS-I.
The upper limits on the WIMP-nucleon spin-independent cross-section are calculated using the
prescriptions of Ref.~\cite{lewin}, and the optimal interval 
method~\cite{yellin} (Fig.~\ref{fig:limits}).
A cross-section of \siglim is excluded at
90\%CL for a WIMP mass of \sigmas.
The sensitivity above 125 GeV/c$^2$ is better than those reported by the experiments
XENON~\cite{xenon} and ZEPLIN~\cite{zeplin}.
It is a factor two away from the cumulative limit reported by CDMS~\cite{cdms}.
The present work represents more than one order of magnitude 
improvement in sensitivity compared
with the previous EDELWEISS results~\cite{edw} based on detectors without
surface event identification.
The sensitivity can be further improved by an increase of exposure,
and, as of end 2009, the present run is being pursued until a significant
number of additional detectors are ready to be installed in the EDELWEISS-II cryostat.

As a further indication of the robustness of the ID technology, it should be noted
that no events were observed in the nuclear recoil band in the exposure of 
18 \kgd of the detector not included in the present search because of its missing
guard  and veto channels.

\section{Conclusion}

The EDELWEISS II collaboration has performed a direct search for
WIMP dark matter using nine 400~g heat-and-ionization cryogenic
detectors equipped with interleaved electrodes for the rejection
of near-surface events.
A total effective exposure of \kgdefftot \kgd
has been obtained after six months of operation in 2009 at the Laboratoire 
Souterrain de Modane and additional data from earlier runs with two  
detectors in 2008.

The observation of one nuclear recoil candidate above 20 keV
is interpreted in terms
of limits on the cross-section of spin-independent interactions
of  WIMPs and nucleons.
Cross-sections of \siglim are excluded at
90\%CL for WIMP masses of \sigmas. 
Further analysis are going on in order to reduce
the effective threshold of the detectors to better address the
case of lower-mass WIMPs.
This result demonstrates for the first time the very high rejection
capabilities of these simple and robust detectors in an actual
WIMP search experiment.
It also establishes the ability of the EDELWEISS-II experiment
for long and stable low-radioactivity data taking and for
the rejection of neutron-induced nuclear recoils.

The EDELWEISS collaboration aims to increase the cumulated 
mass of the detectors in operation at the LSM, 
in order to soon probe the physically significant 10$^{-8}$ pb 
range of spin-independent  WIMP-nucleon cross-sections.

\section{Acknowledgments}
The help of the technical staff of the Laboratoire 
Souterrain de Modane and the participant laboratories is 
gratefully acknowledged. 
The contribution of B. Branlard is particularly recognized.
The Collaboration also recognizes the dedication and efforts
of D. Drain, C. Goldbach and G. Nollez for the 
preparation and installation of EDELWEISS-II.
This project is supported in part by the
Agence Nationale pour la Recherche under contract ANR-06-BLAN-0376-01, 
and by the
Russian Foundation for Basic Research (grant~No.~07-02-00355-a).

\newpage

\begin{thebibliography}{10}

\bibitem{rev} G. Jungman, M. Kamionkowski, and K. Griest, Phys. Rep. 267 (1996) 195; 
G. Bertone, D. Hooper and J. Silk, Phys. Rep. 405 (2005) 279; \\
R. J. Gaitskell, Annual Rev. Nucl. and Part. Sci. 54 (2004) 315.
\bibitem{goodman} M. W. Goodman and E. Witten, Phys. Rev. D 31 (1985) 3059.
\bibitem{cdms} Z. Ahmed et al., Phys. Rev. Lett. 102 (2009) 011301.
\bibitem{xenon} J. Angle et al., Phys. Rev. Lett. 100 (2008) 21303.
\bibitem{zeplin} V.N. Lebedenko et al., Phys. Rev. D 80 (2009) 052010.
\bibitem{edw} V. Sanglard et al., Phys. Rev. D 71 (2005)122002;\\
X.F. Navick, Proc. International Workshop on Low Temperature Detectors - LTD13,
AIP Conf. Proc. 1185 (2009), in press.
\bibitem{brinkbroni} P.L. Brink, et al., Nucl. Instrum. Meth. A 559 (2006) 4148\\
A. Broniatowski, et al., J. Low Temp. Phys. 151 (2008) 830.
\bibitem{id} A. Broniatowski et al., Phys. Lett. B 681 (2009) 305.
\bibitem{edwsetup} Paper in preparation.
\bibitem{eureca} H. Kraus et al., Nucl. Phys. B (Proc. Suppl.) 173 (2007) 168.
\bibitem{lemrani} R. Lemrani et al., J. Phys. Conf. Ser. 39 145 (2006) 145.
\bibitem{markus} M.~Horn,  Ph.D. thesis, scientific report FZKA7391 \\
(http://bibliothek.fzk.de/zb/berichte/FZKA7391.pdf).
\bibitem{martineau} O. Martineau et al., Nucl. Instrum. Meth. A 530 (2004) 426.
\bibitem{broni-simu}A. Broniatowski, Nucl. Instrum. Meth. A 520 (2004) 178. 
\bibitem{lewin} J.D. Lewin and P.F. Smith, Astropart. Phys. 6 (1996) 87.
\bibitem{yellin} S. Yellin, Phys. Rev. D 66 (2002) 032005.
\bibitem{cresst} G. Angloher et al., Astropart. Phys. 31 (2009) 270.
\bibitem{qxe} E. Aprile et al., Phys. Rev. C 79 (2009) 045807;\\
A. Manzur et al. (2009) preprint arXiv:0909.1063.



\end{thebibliography}
\end{document}